# Comparison of parameters of vowel sounds of russian and english languages

Vladimir Fedoseev [1][0009-0006-3520-5358], Anton Konev [2][0000-0002-3222-9956], Alexey Yakimuk [3][0000-0001-9736-7658]

[1] Tomsk State University of Control Systems and Radioelectronics, 40 Lenin Str., Tomsk, 634050, Russia
fedoseev-vi@yandex.ru
[2] Tomsk State University of Control Systems and Radioelectronics, 40 Lenin Str., Tomsk, 634050, Russia
kaa@fb.tusur.ru
[3] Tomsk State University of Control Systems and Radioelectronics, 40 Lenin Str., Tomsk, 634050, Russia
yay@fb.tusur.ru

**Abstract.** In multilingual speech recognition systems, a situation can often arise when the language is not known in advance, but the signal has already been received and is being processed. For such cases, some generalized model is needed that will be able to respond to phonetic differences and, depending on them, correctly recognize speech in the desired language. To build such a model, it is necessary to set the values of phonetic parameters, and then compare similar sounds, establishing significant differences.

**Keywords:** Speech Technologies, Vowel Parameters, Formants.

## 1 Introduction

Speech technologies recognize, analyze and synthesize human voice. Speech imitation, perception of the meaning of phrases, conversion of speech into text, working with voice as a biometric characteristic - all these are different types of speech technologies. This section of computer science is considered to be one of the most complex, because it is at the intersection of several complex disciplines: linguistics, mathematics and programming.

In 2016, 20 percent of searches on smartphones were made using voice. According to the Global Web Index, one in five Internet users also use voice options. Among young people, the figure is even higher: 25 percent of 16-24-year-olds use voice recognition.

The purpose of this work is to analyze the parameters of the audio signal for the vowel sounds of Russian and English, to establish regularities and to work out the method of analysis, on the basis of which the expansion of the capabilities of the program complex will be carried out in the future.

The program complex sr, developed within the framework of group project training of TUSUR, was used during the work. It can determine:

1. the presence of vocalization of sounds;
2. frequency of the main tone;
3. intensity of the main tone;
4. dynamics of change in the frequency of the main tone;
5. deviation of the main tone frequency;
6. dynamics of change in the intensity of the main tone;
7. ratio of the intensity of harmonics to the intensity of the main tone.

More details about this program are written in the corresponding publications [1][2].

What is most important for this work is the ability of this complex to determine numerical values for the first and second formants in the frequency ranges 0-750 and 750-2500 Hz, presenting these values in the form of easy-to-understand graphs.

## 2 Methods

### 2.1 Terms and basics of speech technologies

Sounds result from simple or complex vibrational movements. An example of a complex sound is the oscillation of a string. If you put a string in a vibrating state, you can get the lowest frequency that the string can produce. This frequency is called the fundamental frequency or source frequency.

The fundamental tone frequency (MTF) is the acoustic correlate of the tone, defined as the vibration frequency of the vocal cords. From an acoustic point of view, the BST is the first harmonic of the speech signal. In each speaker, the basic frequency of the basic tone is individual and is determined by the peculiarities of the larynx structure. On average, it ranges from 80 to 210 Hz for male voices and from 150 to 320 Hz for female voices. The frequency of the main tone sets the period of repetition of oscillations [3].

Any object in the state of oscillation oscillates not only with the basic frequency, but also with a frequency two, three, four, etc. times higher than the basic one. These tones are called overtones. Harmonic overtones together with the main tone are called harmonics and form a natural sound order [5].

In acoustics, overtones are sounds included in the spectrum of musical sound; the height of overtones is higher than the main tone (hence the name).

Human vocal cords make similar vibrations.

The volume of air enclosed in a hollow body has its own frequency of oscillation. If you bring another sounding body to this body, the sound will become louder, because the air will begin to vibrate with its own frequency and will amplify the volume of the original sound. This hollow body will be called a resonator. Several resonators can form a system of resonators. In humans, such a system of resonators will be the oral cavity [4].

The natural frequency of a resonator may coincide with the frequency of the fundamental tone or with one of the overtones, amplifying either the fundamental tone or one of the overtones. These amplified frequencies are called formants.

A formant denotes a certain frequency region in which, as a result of resonance, a certain number of harmonics of the tone produced by the vocal cords are amplified, i.e. in the sound spectrum the formant is a rather clearly distinguishable region of amplified frequencies.

In order to characterize a complex sound it is necessary to know about the frequency of the source, the frequency of the harmonics and how the main tone and harmonics relate to each other in terms of intensity. The result is the spectrum of the sound [5].

The origin of speech in humans is possible with the help of the speech apparatus. This apparatus is a union of connected acting organs (throat, lungs, oral cavity with tongue, pharynx, teeth, lips, nasal cavity) that contribute to the formation of voice, capable of regulating it and forming it into meaningful phrases.

A speech sound is the basic unit of oral speech, characterized by a certain strength, pitch, timbre and duration.

Height is directly proportional to the frequency of oscillations. The more vibrations are made per unit of time, the higher the sound. A person is able to distinguish the pitch within the range from 16 to 20,000 Hz.

The duration of a sound is determined by the amount of time at pronunciation [4].

There are two types of sounds - vowels and consonants.

Vowels are sounds pronounced without obstacles on the way of the air jet in the epiglottal cavities. Some languages distinguish a more complex type of vowels - diphthongs, which are divided into descending and ascending.

Consonants are sounds pronounced with obstacles in the way of the air jet in the supraglottic cavities or in the larynx, characterized by muscular tension in the place of the obstacle, the presence of noise and a strong air jet.

Depending on the active organ, the place and method of obstruction formation, the work of the vocal cords, and the position of the soft palate, an articulatory classification of consonants is made.

According to the acoustic classification consonants are divided into sonorous and noisy. Noisy in turn are ringing and deaf. Also, by additional coloring among consonants distinguish soft and hard.

When pronouncing ringing explosive consonants, the vocal tract closes in some area of the mouth cavity. The air is compressed behind the bow, and then sharply released.

The human auditory apparatus has a peculiarity which consists in masking by the most intensive frequencies of neighboring frequencies having less intensity at the same moment of time. In other words, a person hears only the frequency components that are most intense in some of its neighborhood. Since the main feature of a formant is the amplification of the intensity of the spectrum in a certain frequency region, the most intense harmonics of the main tone will be located in this region of the formant.

Vowels and sonorous consonants have formants.

To distinguish vowels traditionally used distinguishing signs, which are associated with the position of formants FI and FII [5].

The following parameters were used for parametric description of stressed vowels:

- sign of vocalization of the segment;
- frequency of the maximum intensity harmonic lying in the frequency range up to 800 Hz;
- frequency of the second most intense harmonic in the frequency range up to 800 Hz;
- frequency of the maximum intensity harmonic lying in the frequency range from 800 to 2300 Hz;
- frequency of the second most intense harmonic lying in the frequency range from 800 to 2400 Hz.

Formants can be used to divide a stream of sounds into vocalized and non-vocalized sections.

Different methods can be used for vocalized and non-vocalized sections.

One of the parameters characterizing vocalized sounds is their formant structure. Transitions between sounds are determined based on the di-dynamics of changes in the corresponding frequencies. Thus, preliminary segmentation of the speech signal is based on analyzing the dynamics of formant frequency changes [5].

Obviously, the accuracy of the extraction of such parameters as the frequency of the main tone, the intensity of harmonics, the dynamics of their changes, etc., directly depends on the initial stage of speech signal processing. The use of the speech signal processing model on the peripheral part of the human auditory system is one of the most promising methods of preprocessing.

### 2.2    Sample and method description

In Russian phonetics, there are six vowel sounds [a], [o], [y], [и], [э] and [ы]. Their sound is influenced by the stressed-unstressed position in a word and other bordering sounds. For all sounds except [ы] there are four variations of co-seminal sounds: hard-hard, soft-hard, hard-soft, soft-soft, soft-soft. For the sound [ы] can be distinguished only variations hard-hard and hard-soft for the reason that when it is pronounced after a soft sound, it turns into [и]. At the same time, the sound [и] softens consonants and cannot stand before hard. It is also worth noting that "hard" in these combinations means not only a hard consonant sound, but also any vowel, as well as the absence of sound, as if the described sound is at the beginning or end of the word. Therefore, we can have such sounds as [и] between hard and hard, as well as [и] between hard and soft - the hard left here is the absence of a sound as such, as, for example, in the word «иволга» or in the simple conjunction «и».

In this paper, for the phonetics of the Russian language, only vowel percussive sounds in all possible positions will be considered, since the non-percussive vowel sounds in different pronunciations can vary greatly in duration and can disappear altogether, so it is very problematic to track and describe them. Percussive sounds are not subject to such a strong distortion, because they are emphasized in the word as a consequence of percussion - they are emphasized, so they are well distinguishable. Thus, 22 sounds will be considered for Russian phonetics. The sample contains 5 male and 5 female speakers.

For English phonetics, 20 vowel sounds are distinguished, among which are short ([i], [u], [ʌ], [ə], [ɔ], [e], [æ]), long ([i:], [u:], [a:], [ɔ:], [ɔ:]) and diphthong-gi ([iə], [uə], [ai], [ɔi], [əu], [eə], [au], [ei]). The pronunciation of these sounds, unlike in Russian, is not affected by the position between neighboring sounds and accent, as it, in fact, already determines what sound will be pronounced. Accordingly, 20 of the above sounds will be considered for English. The sample size is similar: 5 male and 5 female speakers.

The literature provides models for Russian and English sounds (Fig. 1). Based on them, the correspondence of the measurements to the validity will be established. The data will also be compared with each other. The latter will take most of the work, because it is necessary to develop a methodology for working with a particular software toolkit, taking into account its features, in order to further develop an automated recognition system based on the obtained methodology.

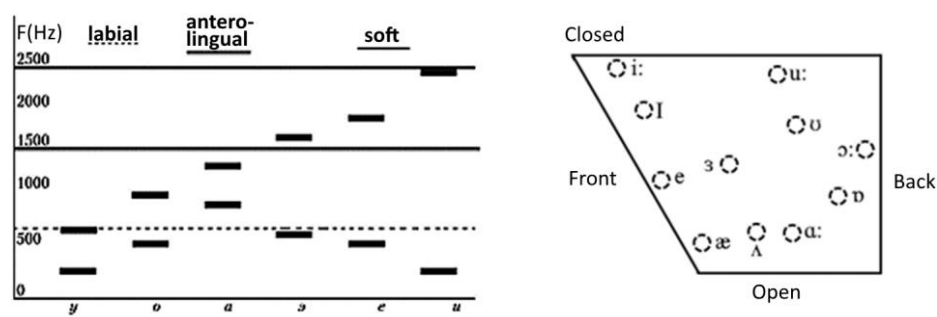

**Fig. 1.** Phonetic models in literature

For comparison of sounds were used harmonic plots or shape-mantle patterns. Signal processing was carried out on the basis of the previously mentioned program complex [1][2]. Here, the first and second harmonics for two frequency intervals: 60-750 Hz and 750-2500 Hz are extracted. For the same or similar sounds, the behavior of these parameters also coincides to a certain extent. Thus, sounds can be studied both independently and in relation to each other.

## 3 Results

In the experimental part, the following patterns were identified (Table 1-2) and comparisons were made (Table 3).

**Table 1.** Regularities of harmonic arrangement for Russian vowel percussion sounds

| Position | Regularity |
|---|---|
| Between two hard | Reduced harmonic values in the upper range. |
| Between hard and soft | Upward character of harmonic values in the upper range. |
| Between soft and hard | Downward character of harmonic values in the upper range. |
| Between two soft | Elevated harmonic values in the upper range. |

**Table 2.** Regularities of harmonic arrangement for Russian vowel percussion sounds

| Sound | Regularity |
|---|---|
| [a] | In lower range 200-750 Hz . First harmonic over the second.<br>In upper range 800-900 Hz for the first harmonic, 1000-1200 Hz for the second. |
| [o] | In lower range 400-600 Hz. First harmonic over the second.<br>In upper range 800-950 Hz for the first harmonic, 1900-2200 Гц для 2-ой гармоники. |
| [и] | In lower range 200-400 Hz. Second harmonic over the first.<br>In upper range 1750-1850 Hz for the first harmonic, 2200-2400 Hz for the second. |
| [ы] | In lower range 200-400 Hz. Second harmonic over the first.<br>In upper range 800-1300 Hz. The graph has a convex upward shape. |
| [y] | In lower range 300-600 Гц. Second harmonic over the first.<br>In upper range 800-1100 Hz. Second harmonic over the first. |
| [э] | In lower range 450-650 Hz. First harmonic over the second.<br>In upper range 1550-1700 Hz for the first harmonic, 2100-2400 Hz for the second. |

**Table 3.** Comparison of parameters of English phonetic sounds (long, short, diphthongs) with Russian phonetic sounds

| English sound | Expected Russian sound | Similarities, differences |
|---|---|---|
| [a:] | [a] | Clear resemblance. |
| [ɔ:] | [o] | Clear resemblance. |
| [i:] | [и] | Clear resemblance. |
| [u:] | [y] | Resemblance unclear. The values in the upper range are overestimated. |
| [ə:] | [o] or [э] | Resemblance to the values obtained, but divergence from the "book model". |
| [i] | [и] | Clear resemblance. |
| [u] | [y] | Resemblance unclear. The values in the lower range are overestimated. |
| [ʌ] | [a] | Resemblance unclear. The harmonics in the lower range are ambiguous. |
| [ə] | [o] | Clear resemblance. |
| [ə] | [э], [o], [a] | No resemblance. |
| [e] | [э] | Resemblance unclear. The harmonics in the lower range are ambiguous and the values are underestimated. |
| [æ] | [э] | Clear resemblance. |
| [iə] | [и] + ([э] or [a]) | Resemblance unclear. The sound looks like a merge of [и] and [э], not a transition from one sound to the other. |
| [uə] | [y] + [э] | Resemblance unclear. The harmonics in the lower range are ambiguous. |
| [ai] | [a] + [и] | Clear resemblance. |
| [ɔi] | [o] + [и] | Clear resemblance. |
| əu] | ([э] or [o]) + [y] | Resemblance unclear. Sound looks more like [o], than [y]. |

# 4     Conclusion

As a result, the values of formants (harmonics) of vowel sounds in Russian and English were analyzed. The regularities of changing characteristics depending on the position of a sound in a word were established, and also the relationship between a sound and its harmonic pattern was established. In the future, it is planned to apply the obtained knowledge and methodology for the development of a software tool for automatic analysis of the sound signal.

## Acknowledgments

This research was funded by the Ministry of Science and Higher Education of Russia, Government Order for 2023–2025, project no. FEWM-2023-0015 (TUSUR).